\documentclass[12pt,a4paper,oneside]{article}
\usepackage{graphicx,amssymb,amsmath,color}
\usepackage[linkcolor={blue},citecolor={red},colorlinks=true]{hyperref}
\usepackage{enumerate}
\usepackage[margin=2cm]{geometry}
\usepackage{slashed}

\def\beq{\begin{equation}}
\def\eeq{\end{equation}}
\def\bea{\begin{eqnarray}}
\def\eea{\end{eqnarray}}

\def\bit{\begin{itemize}}
\def\eit{\end{itemize}}

\def\l{\left}
\def\r{\right}

\def\pr{ {\cal P }}
\def\baa{\begin{array}}
\def\eaa{\end{array}}

\def\sl#1{\mathord{\not\mathrel{{\mathrel{#1}}}}}
\def\d{\partial}

\def\simgt{\mathrel{\lower2.5pt\vbox{\lineskip=0pt\baselineskip=0pt
           \hbox{$>$}\hbox{$\sim$}}}}
\def\simlt{\mathrel{\lower2.5pt\vbox{\lineskip=0pt\baselineskip=0pt
           \hbox{$<$}\hbox{$\sim$}}}}

\def\bfc{\begin{figure}\begin{center}}
\def\efc{\end{center}\end{figure}}
\def\nn{\nonumber\\}

\begin{document}

\begin{flushright}
\hspace{3cm} 
SISSA 247/2020/FISI
\end{flushright}
\vspace{.6cm}
\begin{center}

\hspace{1 cm}{\Large \bf Bubble wall velocity:  heavy physics effects
}\\[0.5cm]

\vspace{1cm}{Aleksandr Azatov$^{a,b,c,1}$,  and Miguel Vanvlasselaer$^{a,b,c,2}$}
\\[7mm]
 {\it \small

$^a$ SISSA International School for Advanced Studies, Via Bonomea 265, 34136, Trieste, Italy\\[0.15cm]
$^b$ INFN - Sezione di Trieste, Via Bonomea 265, 34136, Trieste, Italy\\[0.1cm]
$^c$ IFPU, Institute for Fundamental Physics of the Universe, Via Beirut 2, 34014 Trieste, Italy\\[0.1cm]
 }

\end{center}

\bigskip \bigskip \bigskip

%%%%%%%%%%%%%%%%%%%%%%%%%%%%%%%%%%%%%%%%%%%%%%%%%%%%%%%%%%%%%%%%%%%%%%%%%%
\centerline{\bf Abstract} 
\begin{quote}
We analyse the dynamics of the relativistic bubble expansion during the first order phase transition focusing on the ultra relativistic velocities  $\gamma\gg 1$. We show that fields 
 much heavier than the scale of the phase transition can significantly contribute to the friction and modify the motion of the bubble wall leading to  interesting phenomenological consequences. NLO effects on the friction due to the soft vector field emission are reviewed as well. 
\end{quote}

\vfill
\noindent\line(1,0){188}
{\scriptsize{ \\ E-mail:
\texttt{$^1$\href{mailto:aleksandr.azatov@NOSPAMsissa.it}{aleksandr.azatov@sissa.it}},
\texttt{$^2$\href{miguel.vanvlasselaer@NOSPAMsissa.it}{miguel.vanvlasselaer@sissa.it}}
}}
\newpage

\section{Introduction}
First order phase transitions (FOPT) in the early universe are very interesting phenomena which can lead to a plethora of  cosmological observations, i.e. production of stochastic gravitational wave signals \cite{Witten:1984rs}, matter-antimatter asymmetry \cite{Kuzmin:1985mm,Shaposhnikov:1986jp} or primordial magnetic fields \cite{Grasso:2000wj}. During the FOPT the change of  phase of the system occurs due to the bubble nucleation and it becomes crucial to understand the dynamics of this process. In this paper,
we will focus on the dynamics of the bubble wall expansion and on the friction effects which are induced due to the interaction with the hot plasma (for the previous studies see also 
\cite{Liu:1992tn,Dorsch:2018pat,Konstandin:2014zta,Moore:1995ua,Moore:1995si, Laurent:2020gpg}).

Ideally, in order to answer this question one has to perform the  out-of-equilibrium quantum 
field theory calculation. However in the case of very relativistic bubbles  with a very large
 Lorentz factor,  $\gamma  \gg 1$, a quasi-classical calculation can provide reliable results 
\cite{Bodeker:2009qy,Bodeker:2017cim,Dine:1992wr,Arnold:1993wc,Mancha:2020fzw}. To avoid dealing with complicated quantum out-of-equilibrium effects, in this study, we thus consider only the bubble expansions with $\gamma\gg 1$.
 We will review the results by \cite{Bodeker:2009qy,Bodeker:2017cim} and show that, in the 
presence of new heavy particles, there is an additional unsuppressed contribution to the 
friction which can prevent the runaway behaviour of the bubble, which is the main result 
of this paper. We demonstrate the importance of this effect using a two-scalars toy model with FOPT.
Next we move on to the discussion of the Next-To-Leading Order  (NLO)  friction effects along the lines of \cite{Bodeker:2017cim} and present an alternative derivation using Equivalent Photon Approximation (EPA)\cite{fermi:24,vonWeizsacker:1934nji,Williams:1934ad,landau:34},
 which we believe offers more intuitive  understanding of the friction. 

The manuscript is organized as follows: in the section \ref{sec:transition} we review the LO friction following \cite{Bodeker:2009qy}, then in the section \ref{sec:mixing}, we derive the friction from the heavy particles and provide an example where this can lead to  observational effects. In the section \ref{sec:nlo} we discuss NLO effects and in the section \ref{sec:conc} we finally conclude and summarize our main results.

\section{Transition pressure}
\label{sec:transition}
Let us start by reviewing the origin of the friction effects focusing on the bubbles which are expanding relativistically $\gamma\gg 1$.
Our discussion will follow closely the presentation in \cite{Bodeker:2009qy,Bodeker:2017cim,Dine:1992wr}. Let us suppose that we are looking at the effects coming from a particle $A$ hitting the wall and producing a $X$ final state (which can perfectly be a multiparticle state) (see Fig .\ref{fig:cartoon}), then the pressure will be given by\\
\bea
\pr_{A\to X}=\int \frac{p_z d^3 p}{p_0 (2\pi)^3} f_A(p)\times \sum_X\int dP_{A\to X}(p_A^Z-\sum_{X} p_X^Z),
\eea
where the first factor is just a flux of incoming particles and the second includes the differential probability of the transition from $A$ to $X$, $dP_{A\to X}$, as well as momentum transfer to the wall $(p_A^Z-\sum_{X} p_X^Z)$.
Note that the equation above is valid if only the mean-free-path of the particles is much larger than the width of the wall, so that we can ignore the thermalization effects inside the wall and consider individual particle collisions with the wall \cite{Arnold:1993wc}. 
 The probability of transition can be calculated as follows
\bea
dP_{A\to X}= \prod_{i \in X}\frac{d^3 k_i}{(2\pi)^3 2 k_i^0}\langle \phi|T|X\rangle\langle X|T|\phi \rangle,
\eea
where $\phi$ is the wave-packet building the one-particle normalized state
\bea
\label{eq:def}
&|\phi \rangle= \int \frac{d^3 k}{(2\pi)^3 2 k_0}\phi(k)|k\rangle,~~\langle p|k \rangle=2p_0(2\pi)^3 \delta^3(p-k)\nonumber\\
&\int\frac{d^3 p}{(2\pi)^3 2p_0}|\phi(p)|^2=1 .
\eea
\bfc
\includegraphics[scale=0.5]{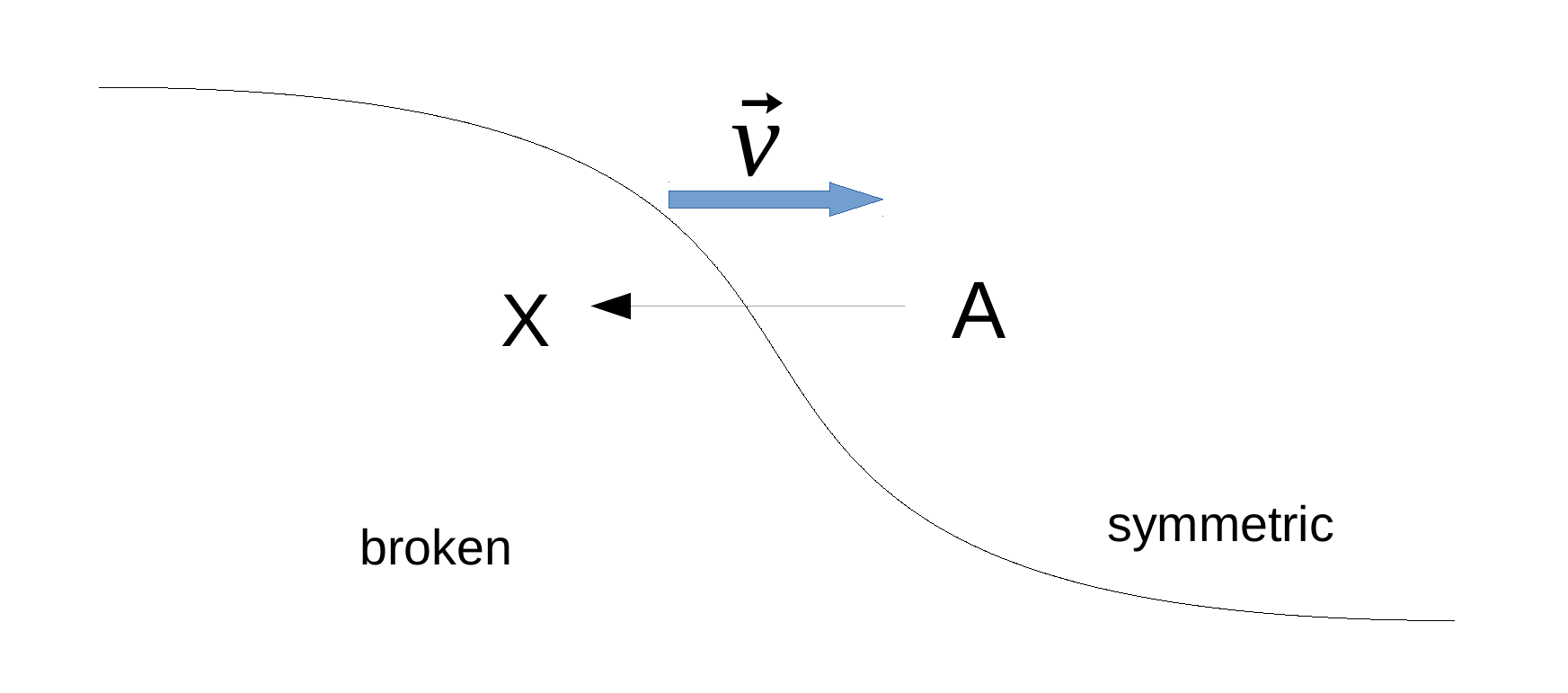}
\caption{Cartoon of a bubble wall  interpolating between the values of the VEV of the scalar field in the \emph{symmetric} and in the \emph{broken} phase.  {The domain wall hitting the plasma in the symmetric phase induces} a $A\to X$ transition. \label{fig:cartoon}}
\efc
Combining all of this and using the energy and transverse momentum conservation we arrive at the following expression for the pressure from $A \to X$ transitions \cite{Bodeker:2017cim}
\bea
\label{eq:master}
\pr_{A\to X}=\int \frac{d^3 p}{(2\pi)^3 2p_0}f_p \prod_{i \in X} \int \frac{d^3k_i}{(2\pi)^3 2 k_0^i}(2\pi)^3\delta^2 (p_\perp-\sum_{i \in X} k_\perp)\delta(p_0-\sum_{i \in X}  k_i^0)| \mathcal{M}|^2(p_A^Z-\sum_{i \in X}  k_i^Z)
\eea
where we have ignored the high density effects for the final particles and $\mathcal{M}$ is defined as follows
\bea
\label{eq:master}
\langle p|H_{\text{int}}|k_1...\rangle  &=& (2\pi)^3\delta^2 (p_\perp-\sum_{i \in X}  k_{\perp,i})\delta(p_0-\sum_{i \in X}  k_i^0)\mathcal{M},\nonumber\\
\mathcal{M} &=& \int dz \chi_p(z)\prod_{i \in X}  \chi_i(z)V .
\eea
Armed with this expression we can proceed to   the calculation of the friction effects. 

\subsection{Leading order (LO) friction}
In this first section, we review the Leading-Order (LO) effects i.e. when the initial and the final state contain one particle ($1\to 1$  transition). We will be focusing on the very relativistic bubble expansions and, in particular, on regimes where the \emph{WKB approximation} is valid, which is when
\bea
p_z L\gg 1,
\label{WKB_app}
\eea
where $L$ is a typical width of the wall and $p_z$ is the momentum of the incident particle.
Let us suppose that the mass $m_1$, in the symmetric phase, changes when passing through the wall to $m_2$ in the broken phase, $m_1 \to m_2$.  According to Eq.\eqref{eq:def}, the matrix element for this transition is equal to
\bea
\langle p|k \rangle=2p_0(2\pi)^3 \delta^3(p-k)\Rightarrow  \mathcal{M}_{1\to 1}=2p_0.
\eea
Then the pressure for the relativistic particles is equal to:
\bea
\mathcal{P}_{1\to 1}=\int \frac{d^3 p}{(2\pi)^3 }f_p (p_s^z-p_h^z)\simeq \int \frac{d^3 p}{(2\pi)^3 }f_p \times \frac{m_2^2-m_1^2}{2 p_0} = \int \frac{d^3 p}{(2\pi)^3 }f_p \times \frac{\Delta m^2}{2 p_0}
\eea
where we have expanded the momenta in $m_{1,2}^2/p_0^2$  and  defined $\Delta m^2 \equiv m_2^2-m_1^2$. { $p_s$($p_h$) denotes the momentum of the particles on the symmetric (broken) side of the bubble wall.}

 It is well known that the quantity $\frac{d^3 p}{p_0}$ is invariant under boost, which allowed the authors of \cite{Bodeker:2009qy}  to conclude that the Leading-Order friction is independent of $\gamma$ and  scales as 
\bea
\label{eq:LOBM}
\mathcal{P}_{1\to 1}\simeq \frac{\Delta m^2 T^2}{24}.
\eea
Let us make a few comments about this result. For simplicity, let us assume first that $m_1=0$
then  it is obvious that the value in Eq. \eqref{eq:LOBM} will be reached only for the   $\gamma$ factors satisfying 
\bea
{\gamma \gtrsim \frac{m_2}{T}},
\eea 
otherwise initial particles simply will not have enough energy to pass through the wall (see more details in the Appendix \ref{sec:pressuredetails} as well as \cite{Mancha:2020fzw} for analytical results).  
 Now let us look at the scenario 
when the initial mass as well is non-zero, $m_1\neq 0$.  In this case the particle will contribute only  if its mass is smaller than the temperature
\bea
m_1 \lesssim T
\eea
otherwise  the contribution of this particle to the pressure will be exponentially suppressed by a Boltzmann factor (see Appendix \ref{sec:pressuredetails}).

{
At last let us comment on the recent papers \cite{Mancha:2020fzw,Balaji:2020yrx}  which have calculated the friction assuming  \emph{local thermal equilibrium} (LTE) around the wall and where a friction scaling like $\propto \gamma^2$ was found. We would like to stress that our calculation applies only to the ultrarelativistic bubbles where LTE is not valid any more,
since the diffusion process is efficient only for moderate velocities,  as was emphasized in
\cite{Mancha:2020fzw}. Even more in the case of ultrarelativistic bubbles  the Ref.  \cite{Mancha:2020fzw} finds a saturation of the friction ( this limit is dubbed ``ballistic''), which is  in agreement with the  results of \cite{Bodeker:2009qy} (i.e. Eq.\eqref{eq:LOBM}).}

\section{Friction from mixing}
\label{sec:mixing}
In the context of a very relativistic bubble, in the rest frame of the wall, the particles colliding it can reach very high energies $\sim \gamma T_{\text{nuc}} $, much larger than the temperature of the transition $\sim T_{\text{nuc}}$ and the symmetry breaking parameter $\sim \langle s \rangle $. In such circumstances, it becomes interesting to investigate whether new degrees of freedom, absent in the low energy lagrangian describing the phase transition, can play a role in the dynamics of the bubble acceleration. The simplest example  where this phenomena can occur is the following:
let us consider the lagrangian of a \emph{massless} fermion mixed with another \emph{heavy} vectorlike fermion
\bea
{\cal L}_{\text{fermion}}= i \bar \psi \sl \d \psi+i\bar N \sl \d N+M \bar N N + Y_{\text{mixing}}  s \bar \psi N 
\label{heavy_lag}
\eea
where $s$ is a field developing a VEV via the phase transition, $\psi$ is the light fermion and $N$ is the heavy fermion. In the regime $M\gg \langle s \rangle  \sim T_{\text{nuc}}$, then at the temperature of the transition, the species $N$ can be ignored (they are Boltzmann suppressed and are not part of the plasma), so their contribution to the pressure naïvely should be zero. However let us consider a process of $\psi$ hitting the wall.
We suppose that the energy of the incident $\psi$ particles is much larger than the mass of the heavy species $N$; $E\gg M\Rightarrow \gamma T\gg M$. 
Note that the mass eigenstates inside and outside of the bubble are different due to the VEV of the $\langle s \rangle $ and in particular there will be mixing between the $\psi$ field and the heaviest mass eigenstate in the broken phase. The mixing angle $\theta_{\psi N}$ is approximately given 
\bea
\sin \theta_{\psi N}\sim \frac{Y_{\text{mixing}} \langle s \rangle  }{M}.
\label{mix_angle}
\eea
 { From this mixing angle, we can deduce that, if the transition is satisfying the condition  $\Delta p_z L \ll 1$,   (where $L$ is the width of the wall)
 there will be a probability of transition  $\psi \to N$ of the form 
\bea
P(\psi \to N)\sim \sin^2 \theta_{\psi N}\sim \frac{Y_{\text{mixing}}^2  \langle s \rangle ^2 }{M^2}.
\label{eq:prob_trans}
\eea
Using the terminology of the neutrino oscillation in matter \cite{Mikheev:1986if} (see for review\cite{Bilenky:2010zza}), to find an unsuppressed transition, we need to be in the regime of \emph{non-adiabaticity}.  In the opposite case,  when the 
transition is \emph{adiabatic} and  $\Delta p_z L \gg 1$ 
is satisfied, the
 incoming $\psi$ remains in the lightest mass eigenstate
 so that the process $\psi\to N$ will be strongly suppressed\footnote{We thank the referee of JCAP for emphasizing this effect to us.}. 
Intuitively the adiabatic regime corresponds to the situation when the evolution of the mixing parameter is so slow that system can instantaneously adapt and the state remains in the same energy level.
We can estimate the pressure due to this mixing using the results of the previous section. We obtain
\bea
\label{eq:fricapp}
\mathcal{P}_{\text{mixing}} &\sim & \underbrace{\int \frac{d^3 p}{(2\pi)^3}f_p}_{\text{Incident $\psi$ density}} \underbrace{P(\psi \to N)}_{\text{Probability of transition}}\times \underbrace{\frac{M^2}{2E}}_{\text{momentum transfer}}\nn
& \sim &  Y_{\text{mixing}}^2  \langle s\rangle^2 \int \frac{d^3 p}{(2\pi)^3 E}f_p\theta (E-M) \theta \l(E-M^2 L\r)\nn
&\sim &Y_{\text{mixing}}^2  \langle s\rangle^2  T^2 \theta(\gamma T-M ) \theta(\gamma T-M^2 L )
\\ \nonumber
&=&Y_{\text{mixing}}^2  \langle s\rangle^2  T^2\theta(\gamma T-M^2 L ),
\eea
where the $\theta$-functions have appeared in order to impose that the initial particle is energetic enough to produce the heavy state and that the process is non-adiabatic, we have also assumed that $M\gg L^{-1}$, so that the first $\theta$ function  becomes redundant.
 In the next section, we will derive more accurately the condition of non-adiabaticity, by explicitly looking at the transitions with a finite wall width.
Note that this new contribution to the pressure is not suppressed by the large $M$ mass and can be present even if $M\gg \langle s \rangle $.}

\subsection{Friction from mixing: more details}
\label{sec;mixing_more_details}
One can derive the expression for the friction force in Eq.\eqref{eq:fricapp} using the master equation Eq.\eqref{eq:master} for the  $\psi \to N$ transitions. Indeed, in  
 the WKB approximation, the solutions for the wave functions  are given by 
\bea
\chi(z)\simeq \sqrt{\frac{k_{z,s}}{k_z(z)}}\exp \l(i \int_0^z k_z(z')d z'\r),
\eea
where $k_{z,s}$ is the $z$ component of the momenta on the symmetric side of the wall.
{ As we have seen above in the adiabaticity discussion, the wall width should play a role in the computation of the pressure effect. As an illustration, let us approximate the wall using a \emph{linear}  ansatz  
\bea
 \langle s\rangle =\l\{\baa{c} 0,~~z <0\\
 v_s\frac{z}{L}  ~~~0\leq z\leq L\\
 v_s ~~~z>L
 \eaa\r. .
  \eea
In the notations of \cite{Bodeker:2017cim}, the $\mathcal{M}$ matrix takes the form  
  \bea
  \mathcal{M} = \int\limits_{-\infty}^{\infty} dz e^{i \int\limits_0^{z} p_{z}^{\psi}(z') dz'} e^{-i \int\limits_0^{z} p_{z}^{N}(z') dz'} V(z) \approx \int\limits_{-\infty}^{\infty} dz e^{i  p_{z}^{\psi} z} e^{-i  p_{z}^{N} z} V(z) = \int\limits_{-\infty}^{\infty} dz e^{i \Delta p_z z} V(z)
  \eea
  where we defined $\Delta p_z \equiv p_{z}^{\psi, in}- p_{z}^{N, out} $, the difference of momentum, and we safely neglected the change of momentum  due to the modifications of the masses induced by the VEV, since these effects are largely subdominant 
 with respect to the heavy mass $M$,
\bea
\frac{Y_{\rm mixing}^2 \langle s\rangle^2}{ E}\ll \frac{M^2}{E}.
\eea
 \emph{Energy} and \emph{transverse momentum} conservation dictate that $p_z^{\psi,N}$  are different on the two sides of the wall, leading to an effective non-zero $\Delta p_z$. 
  Performing the integral, the matrix element splits into three different pieces
   \bea
   \label{eq:wkb11}
 \mathcal{M}& =& \mathcal{M}_{\text{before wall}}+ \mathcal{M}_{\text{inside wall}}+ \mathcal{M}_{\text{after wall}}
 \\ \nonumber
 &=& V_h\frac{1 - e^{i\Delta p_z L}}{iL \Delta p_z^2} + \frac{V_s}{i\Delta p_z},
 \eea
 where the $h,s$ subscripts denote the interactions and momenta inside and outside of the bubble.
The process under study is only possible on the broken side of the wall and thus $V_s = 0, \quad V_h\neq 0$. We obtain
\bea
|V_h|^2=2 Y_{\text{mixing}}^2  \langle s \rangle ^2 p_z^\psi\Delta p_z.
\label{eq:Vh11}
\eea
%On top of this there will be an effect due to the mass modification of $\psi$ and $N$, however this effect will be subleading and suppressed by additional  powers of $Y_{\text{mixing}}^2  \langle s \rangle ^2/M^2$.
Combining the  expression for the matrix element Eq.\eqref{eq:wkb11} with the expression for the vertex Eq.\eqref{eq:Vh11}, the matrix element squared becomes
\bea
 |\mathcal{M}|^2 \approx \frac{V_h^2}{\Delta p_z^2} \bigg(\frac{\sin \alpha}{\alpha} \bigg)^2 = 2\frac{\langle s \rangle ^2 p_z^\psi}{\Delta p_z} \bigg(\frac{\sin \alpha}{\alpha} \bigg)^2 Y_{\text{mixing}}^2  , \qquad \alpha = \frac{L \Delta p_z}{2} \approx \frac{M^2 L}{4E},
 \eea
 with $\Delta p_z = p_z^\psi-\sqrt{(p_z^\psi)^2-M_N^2}$.  
  Plugging it in the master Eq.\eqref{eq:master}, we obtain the following estimate for the mixing pressure\footnote{The integral below is assumed to be taken for the values of $p_z> M_N$, otherwise the process is forbidden.}
\bea
\mathcal{P}_{\text{mixing}}=\int \frac{d^3 p}{(2\pi)^3} f_p \times \frac{ Y^2_{\rm mixing}  \langle s\rangle^2}{2\sqrt{p_z^2-M_N^2}} \times  \bigg(\frac{\sin \alpha}{\alpha} \bigg)^2.
\label{Pres_mixing_WKB}
\eea
As $\big(\frac{\sin \alpha}{\alpha} \big)^2  \to 0$ for $\alpha \gg 1$, we can see that the pressure is suppressed for $\alpha \gg 1$, so that the suppression regime is given by
\bea
\frac{M^2 L}{E}\gg 1 \quad \Rightarrow \quad E\ll M^2 L\sim \frac{M^2}{\langle s \rangle},
\eea
where we have assumed that the width of the wall  $L$, in the wall frame scales like the inverse of the VEV.  Using the fact that the energy of the incident particles is approximately equal to $\sim \gamma T$ we obtain a necessary constraint on the masses of the heavy particles which can be produced
 \bea
 \gamma T\gtrsim \frac{M^2}{\langle s\rangle},~~~M\lesssim \sqrt{\gamma T \langle s\rangle }
 \label{condition_short}
 \eea
 which exactly corresponds to the regime where  the passage of the particle through the wall cannot be treated adiabatically. We can see also that, in the limits $\alpha \ll 1$ and $\alpha \gg 1$, the suppression factor of Eq.\eqref{Pres_mixing_WKB} behaves effectively as a $\theta$-function, reducing  to the estimate in Eq.\eqref{eq:fricapp}. }

We would like to emphasize that the pressure from the mixing is not suppressed by the mass of the heavy particles and in general can be present even  if we treat our theory as an effective field theory (EFT) with heavy degrees of freedom integrated out.

 One can also ask what could be the maximal pressure from the mixing in this case. We can estimate it by using unitarity arguments on the maximal value of the mixing  coupling $Y^{\rm max}_{\text{mixing}}\sim 4 \pi$.  So that, the maximal pressure from mixing is
\bea
\mathcal{P}_{\text{mixing}}^{\text{max}}\simeq \frac{T^2}{48}(16\pi^2) \langle s \rangle ^2\theta(\gamma T -M^2L). 
\eea
In the appendix \ref{sec:frictionheavy}, we give other examples  of friction induced by the otherwise decoupled particles in the theories with only scalars.

\subsection{Importance of friction from heavy particles}
One can wonder whether this friction from mixing can be 
phenomenologically important, since in any case we are looking at the very 
relativistic bubble expansion {velocities}   $v\to 1$. However, it is known that 
relativistic bubbles in runaway regime have all their energy stored in the wall kinetic motion, while bubbles which have reached a terminal velocity have vanishingly 
small fraction of energy stored in the wall and  most of the energy released in the phase transition is transferred to the sound waves (plasma motion) \cite{Ellis:2018mja,Ellis:2019oqb,Espinosa:2010hh}. 
This different distribution of energy has important phenomenological consequences
on the spectrum of stochastic gravitational wave background  since the \emph{bubble wall collisions} signal $\Omega_{\phi}$ and \emph{plasma  motion} signal $\Omega_{sw}$ lead to different shape of the spectrum (see for example \cite{Caprini:2019egz}). Namely, the most obvious difference is the fall of the signal at high frequencies;
\bea
\Omega_{sw, f \to \infty}  \propto f^{-4} \text{~ (terminal velocity)}, \qquad \Omega_{\phi, f \to \infty}  \propto f^{-3/2} \text{~ (Runaway)}.
\eea
{
In order to understand whether the friction from mixing can indeed prevent the runaway bubble case, let us consider the following toy model (\cite{Breitbach:2018ddu,Azatov:2019png}) described in the infrared region by the lagrangian
\bea
\label{eq:scalarsimp}
{\cal L}_{IR}&=&\frac{1}{2}(\d_\mu \phi)^2+\frac{1}{2}(\d_\mu \eta)^2-\frac{m^2_{\phi} \phi^2}{2}-\frac{m^2_\eta \eta^2}{2}-\frac{\lambda_\phi}{4} \phi^4-\frac{\lambda_\eta}{4} \eta^4- \frac{\lambda_{\phi \eta}}{2} \phi^2 \eta^2+i\bar \psi \sl \d \psi .
\eea
We will assume that at high energies this lagrangian is UV completed to
\bea
{\cal L}_{UV}&=&\frac{1}{2}(\d_\mu \phi)^2+\frac{1}{2}(\d_\mu \eta)^2-\frac{\tilde m^2_{\phi}\phi^2}{2}-\frac{\tilde m^2_\eta \eta^2}{2}-\frac{\tilde\lambda_\phi}{4} \phi^4-\frac{\tilde \lambda_\eta}{4} \eta^4- \frac{\tilde\lambda_{\phi \eta}}{2} \phi^2 \eta^2\nn
&&+ i \bar \psi \sl\d \psi+ i \bar N \sl\d N-M \bar N N-(Y_{\rm mixing } \bar \psi \phi N+h.c.), 
\eea
where all of the parameters $\tilde m^2_{\phi,\eta}, \tilde \lambda_{\phi,\eta,\phi\eta}$  are the parameters of UV theory and 
$m^2,\lambda$ in the Eq.\eqref{eq:scalarsimp} are the parameters of IR effective theory obtained by matching after integrating out the heavy fermion $N$. 
We will assume that IR lagrangian is defined at the scale of the symmetry breaking of the theory, which is much smaller than the mass of the fermion $N$, $ \langle s\rangle, m_\phi,m_\eta \ll M$.
This introduces the usual  tuning into the model, which is analogous of the Higgs boson hierarchy problem in the presence of heavy new physics.  However we will not bother about a solution to this hierarchy  problem and take  Eq.\eqref{eq:scalarsimp} as a toy, very fine-tuned example to illustrate  the effect of the  friction from the mixing.
}

Let us consider the potential for the scalar fields of the model.
On the top of the tree-level potential, at one loop, the usual Coleman-Weinberg potential is generated \cite{Weinberg:1973am} for the fields $\phi,\eta$  (in $   \overline{MS}$ scheme) (we are using the IR lagrangian of Eq.\eqref{eq:scalarsimp})
\bea
V_{CW}=\sum_{i = \eta,\phi} \frac{m_i^4}{64 \pi^2}\l[\log \frac{m_i^2}{\mu_R^2}-\frac{3}{2} \r].
\eea
The thermal corrections can be taken into account by adding the thermal potential 
 $V_T$ defined as follows
\bea 
V_T=\sum_{i = \eta,\phi}\frac{T^4}{2\pi^2} J\l(\frac{m_i^2}{T^2}\r),~~~J(y^2)\equiv \int_0^\infty dx x^2 \log \Big[1-\exp\big(-\sqrt{x^2+y^2}\big)\Big].
\eea
Higher loop corrections  due to the daisy diagrams can be taken into account using the truncated full dressing procedure\cite{Curtin:2016urg} 
 \begin{equation}
  V(\phi,\eta,T)=V_{tree}(\phi,\eta)+\sum_{i = \eta,\phi} V_{CW}(m_i^2+\Pi_i^2)+V_T(m_i^2+\Pi_i^2).
 \end{equation}
In the case of the model \eqref{eq:scalarsimp}, the thermal mass corrections are given by
\bea
 & m_\phi^2+\Pi^2_{\phi}= m_\phi^2+3\lambda_\phi \phi^2+\lambda_{\phi \eta}\eta^2+T^2\l(\frac{\lambda_\phi}{4}+\frac{\lambda_{\phi \eta}}{12}\r),\\
 & m_\eta^2+\Pi^2_{\eta}= m_\eta^2+\lambda_{\phi \eta}\phi^2+3\lambda_\eta \eta^2+T^2\l(\frac{\lambda_\eta}{4}+\frac{\lambda_{\phi \eta}}{12}\r).
 \eea
{
Generically we have to analyze the phase transition in the $(\phi,\eta)$ field  space, 
however the discussion simplifies  if we put the coupling $\lambda_\phi=m_\phi=0$.
Indeed in this case, along the line $\eta=0$, the tree-level potential is vanishing 
and only the one loop potential will be controlling the phase transition. The tree-level $\eta^4$-potential is stabilizing the $\eta$-direction, thus  it is obvious that the tunnelling must happen along $\eta=0$ direction.  The calculation becomes even simpler if we set $m_\eta=0$
 then the  only mass parameter in this construction  is the renormalization scale $\mu_R\equiv \lambda_{\phi\eta} w$ which  fixes the value of the VEV of the field $\langle \phi \rangle \sim w$.}

The transition  from the false to the true vacuum, separated by the potential barrier, can be calculated  using the 
usual \emph{bounce action}  (see \cite{Coleman:1977py,Linde:1980tt,Linde:1981zj})

\bea
\Gamma(T)\sim \text{max} \l[T^4 \l(\frac{S_3}{2\pi T}\r)^{3/2}\text{Exp}(-S_3/T), R_0^{-4} \l(\frac{S_4}{2\pi}\r)^2 \text{Exp}(-S_4)\r]. 
\eea
However the phase transition in the early universe will occur when the rate of transition becomes comparable to the expansion rate of the universe. This condition defines the \emph{nucleation temperature} $T_{\rm nuc}$ and is approximately given by
\bea
\label{eq:nucleation}
&&\Gamma(T_{\rm nuc})= H^4(T_{\rm nuc}),\nonumber\\
&&H^2\equiv \frac{\rho_{\text{rad}}+\rho_{\text{vac}}}{3 M_{\text{pl}}^2}= \frac{1}{3 M_{\text{pl}}^2}\l(\frac{\pi^2 g_*}{30}T^4+\Delta V \r),
\eea
where $M_{\text{pl}} \equiv 2.435 \times 10^{18}$ GeV is the reduced Planck mass. 

We are prepared now to discuss the friction effects. The bubble will have runaway behaviour
 if the LO friction,  which in our model is equal to 
\bea
\mathcal{P}_{{\rm{LO}}}\simeq  \frac{T_{\rm nuc}^2}{24}\lambda_{\phi \eta} \langle \phi \rangle ^2 \theta (\gamma T-\langle \phi \rangle  \sqrt{\lambda_{\phi\eta}}),
\eea
cannot overcome the potential difference, providing the driving force for the expansion of the bubble. This amounts to the condition \cite{Espinosa:2010hh}
\bea
\Delta V > \mathcal{P}_{\rm LO} ~~\text{~(runaway condition).}
\label{eq:runawaycondition}
\eea
At the same time, as we have seen, there can be an additional friction induced by the mixing effect 
{
\bea
\mathcal{P}_{\rm mixing}\simeq \frac{T_{ \rm nuc}^2}{48}Y_{ \rm mixing}^2 \langle \phi\rangle^2 \theta(\gamma T - M^2L),
\eea}
which can prevent the runaway behaviour. Now if the condition 
\bea
\label{eq: mixingeffect}
\mathcal{P}_{\rm LO}+ \mathcal{P}_{\rm mixing}>\Delta V > \mathcal{P}_{\rm LO}
\eea
is satisfied we are in the situation when the mixing pressure is preventing the bubbles from the otherwise runaway motion. 
To analyse the $\pr_{\rm mixing}$ effects in our model we have deferred from performing the full parameter scan and instead have fixed the symmetry breaking scale to be $10^5$ GeV and
the mixing coupling $Y_{ \rm mixing}=\sqrt{2}$. Then the  region of the parameter space  where the mixing effect is important is displayed on the Fig.\ref{fig:couprange}.
\bfc
\includegraphics[scale=0.47]{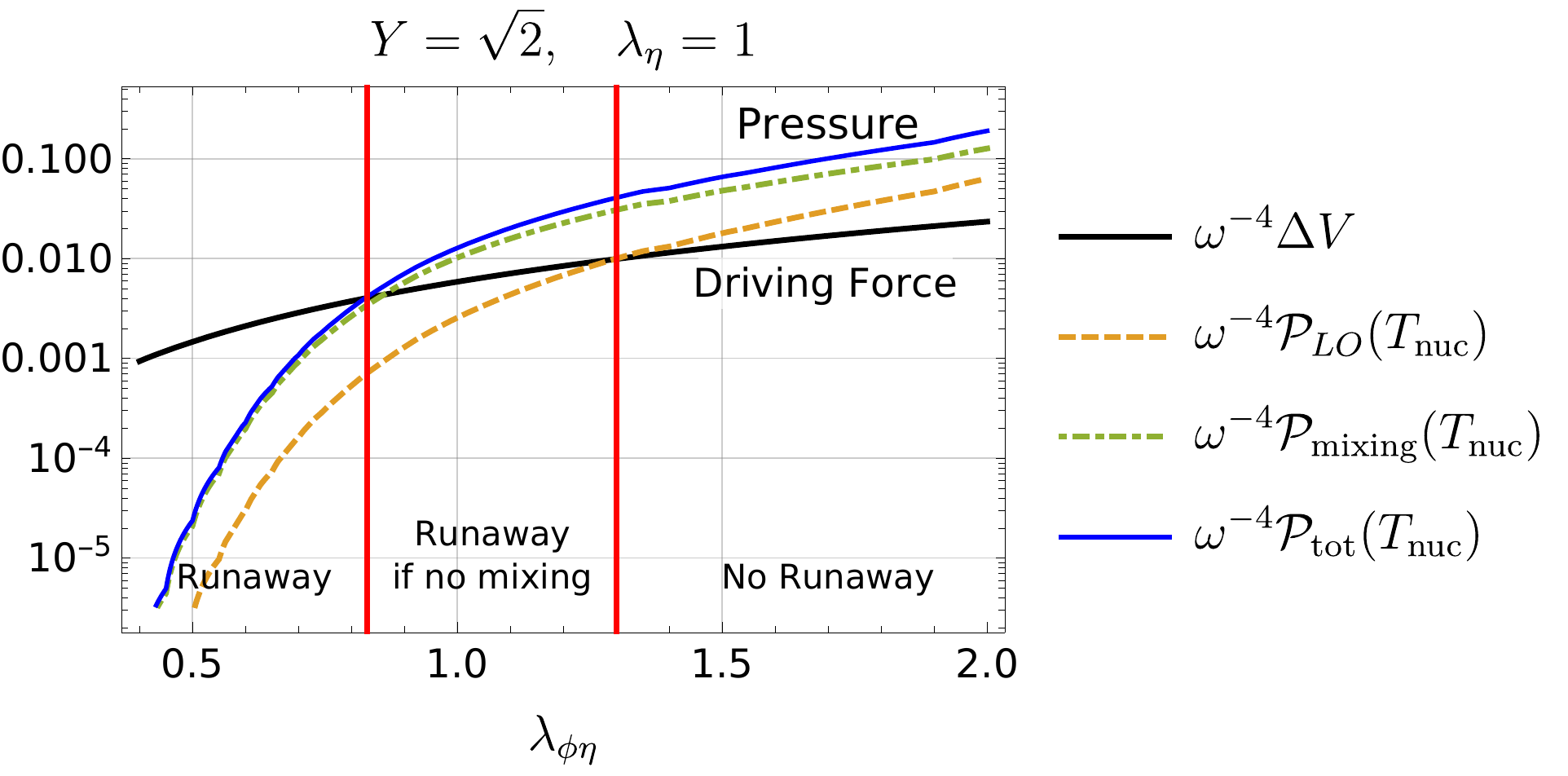}
\includegraphics[scale=0.47]{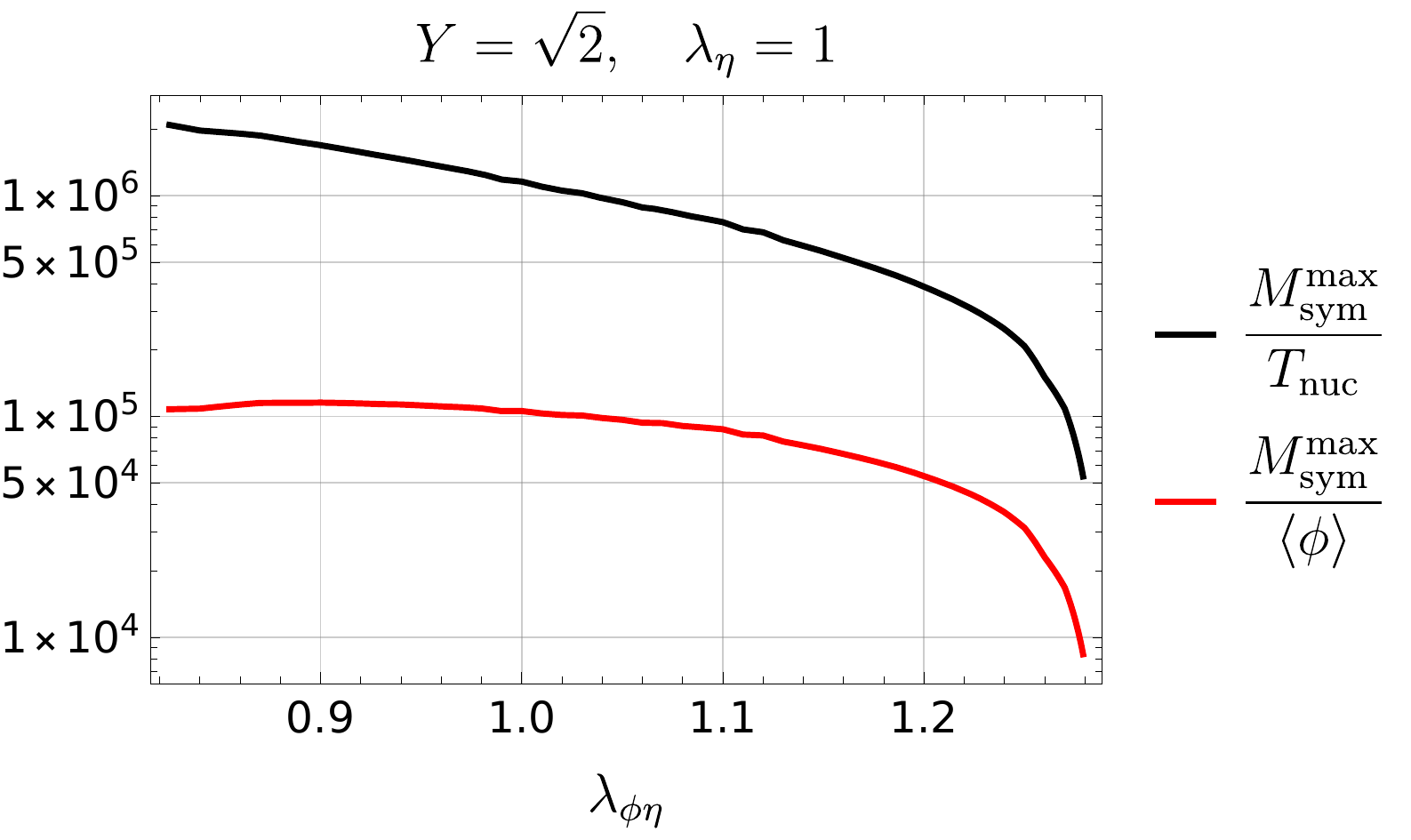}
\caption{\label{fig:couprange}
Left- the potential difference and various contributions to the pressure as a function of the coupling $\lambda_{\phi\eta}$. The scale of the symmetry breaking was fixed to be $w=10^5$ GeV so that $\langle \phi \rangle\sim 10^5$ GeV. Right- The maximal mass of the heavy particle defined by the Eq.\ref{eq:mmax} as a function of  $\lambda_{\phi\eta}$.  As a typical width of the wall, we considered $L \sim 1/ \langle \phi\rangle$. }
\efc

In order to find the upper bound on the masses of the states 
 which can be produced in mixing we need to estimate the maximum value of the Lorentz $\gamma_{\rm max} $ factor that would have been reached if the bubbles kept accelerating till the moment of the collision. It can be estimated   from the ratio of initial and final radii of the bubble and is approximately  equal to 
\cite{Ellis:2019oqb,Azatov:2019png}
\bea
\gamma_{\rm max} \simeq \frac{2 R_*}{3 R_0}\l(1-\frac{\mathcal{P}_{\rm LO}}{\Delta V}\r).
\eea
The initial bubble radius and the bounce solution can be found numerically, while the final radius can be estimated   according to \cite{Enqvist:1991xw} by the derivative of the bounce action:
\bea
R_*=\frac{(8 \pi)^{1/3}}{\beta(T_{\rm nuc})},~~~\beta(T)=H T\frac{d}{dT}\l(\frac{S_3}{T}\r),
\eea
where $H$ is the Hubble constant.
{{
Then the friction from mixing can be generated only by the states satisfying
\bea
\label{eq:mmax}
M<M_{\rm max}= \text{min}\bigg[\gamma_{ \rm max} T_{\rm nuc},  \sqrt{\gamma_{ \rm max}\frac{T_{\text{nuc}}}{L}} \bigg]. 
\eea
For a wall length $L \sim 1/\langle \phi \rangle$, the condition simplifies to 
\bea
\label{eq:mmax_2}
M<M_{\rm max}=  \sqrt{\gamma_{ \rm max} T_{\rm nuc}\langle \phi \rangle}. 
\eea
We report the value $M_{ \rm max}$ on the Fig.\ref{fig:couprange}.
In our model, we can see that states as heavy as $10^{10}$ GeV, $10^{5}$ times heavier than the scale of the transition, can lead to non-vanishing friction effects. Generically, one can  estimate $M_{\rm max}$ as follows
\bea
M_{\rm max}=\sqrt{\gamma_{ \rm max} T_{\rm nuc}\langle \phi \rangle}\sim \sqrt{\frac{R_*}{R_0} T_{\rm nuc}\langle \phi \rangle}.
\eea
The values of the initial and the final radii are very roughly equal to:
\bea
R_0\sim \frac{1}{T_{\rm nuc}},~~R_*\sim H^{-1}\sim \frac{M_{\text{pl}}}{{\rm scale}^2},
\eea
where ``scale" refers to the energy scale of the potential $\sim \langle \phi \rangle$.
Combining all of this together we can find the estimate for the maximal mass to be
\bea
M_{\rm max}\sim T_{\rm nuc}  \l(\frac{M_{\text{pl}}}{\rm scale}\r)^{1/2}.
\eea
Of course this estimate is valid  only for the theories where the bubbles are runaway without the friction from mixing.}}

 In the next section  however  we will review the NLO effects from the gauge field which generically prevent bubbles from infinite acceleration.

\section{NLO effects (review of \cite{Bodeker:2017cim})}
\label{sec:nlo}
So far we have been looking at the  effects appearing in $1\to 1$ transition, now let us move to the $1\to 2$ transitions (we closely follow the discussion in \cite{Bodeker:2017cim}) . Again we will assume that we are in the regime where the WKB approximation is valid i.e. $p_z L\gg 1$, where $L$ is the width of the wall. The calculation of the $1\to 2$ splitting simplifies in the limit when $k_z\gg m,k_\perp $ and, in this case, it becomes easy to find the solution for the free wave functions
\bea
\chi(z)\simeq \sqrt{\frac{k_{z,s}}{k_z(z)}}\exp \l(i \int_0^z k_z(z')d z'\r).
\eea
Using the following notation for the initial and final momenta
\bea
p&=&(p_0,0,0,\sqrt{p_0^2-m_A^2(z)})\nn
k^{(1)}&=&(p_0(1-x),0,k_\perp, \sqrt{p_0^2(1-x)^2-k_\perp^2-m_C^2(z)})\nn
k^{(2)}&=&(p_0 x,0,-k_\perp,\sqrt{p_0^2 x^2-k_\perp^2- m_B^2(z)}),
\eea
 the product of three wave functions in $1\to 2$ splitting is equal to
\bea
\chi_A(z)\chi_B^*(z)\chi_C^*(z)\sim \exp \l[\int_0^z  \l( \frac{m_A^2(z)}{2 p_0}-\frac{m_B^2(z)+k_\perp^2}{2 k^{(1)}_0}-\frac{m_C^2(z)+k_\perp^2}{2 k^{(2)}_0}\r) \r].
\eea
Then the matrix element is equal to
\bea
&&\mathcal{M}= V_s \int_{-\infty}^0 \exp\l[i z \frac{A_s}{p_0}\r]+V_h \int^{\infty}_0 \exp\l[i z \frac{A_h}{p_0}\r]
=2 i p_0 \l(\frac{V_h}{A_h}-\frac{V_s}{A_s}\r)\nn
&&A=-\frac{k_\perp^2}{x(1-x)}+m_A^2-\frac{m_C^2}{1-x}-\frac{m_B^2}{x},
\eea
and the matrix element squared becomes
\bea
\label{eq:matrixel}
|\mathcal{M}|^2=4 p_0^2 \l|\frac{V_h}{A_h}-\frac{V_s}{A_s}\r|^2.
\eea
The reference \cite{Bodeker:2017cim} has studied  various splitting effects and it was  
shown that only the production of the vector particles gaining the mass during the phase 
transition can lead to a friction effect growing with the Lorentz factor $\gamma$.  Let us 
apply this formalism for the case of the QED-like theory. In other words let us consider the 
process $\psi\to A\psi$, where the fermion splits into a vector boson and the fermion. This 
process, which is obviously forbidden by momentum conservation in the absence of the wall, can happen when the wall is present. For the production of transversely polarized vector bosons during transition $\psi \to \psi A^T$, the matrix element takes the form 
\bea
&V_h=V_s=\frac{\sqrt 2 k_\perp}{x},\nn
&|\mathcal{M}_V|^2=\frac{8 p_0^2 k_\perp^2}{x^2} \l|\frac{A_h-A_s}{A_h A_s}\r|^2=\frac{8 p_0^2 m_V^4}{(k_\perp^2+m_V^2)^2 k_\perp^2}.
\label{eq:matrix}
\eea
Focusing on the limit  $k_\perp\sim m \ll k_0\sim k_z$, we recover the following expression for the pressure
 \bea
\label{eq:BM} 
\mathcal{P}_{\psi \to A\psi}\simeq\int\frac{d^3 p}{8 p_0^2 (2\pi)^6} f_p\int \frac{d k_0^{(2)}}{k_0^{(2)}}\int d^2 k_\perp|\mathcal{M}|^2 \frac{k_\perp^2+m_V^2}{2 p_0 x}.
 \eea
Plugging in our expression for the matrix element \eqref{eq:matrix} we get:
\bea
\label{eq:pres12}
\mathcal{P}_{\psi \to A\psi}&\simeq&\int\frac{d^3 p}{8 p_0^2 (2\pi)^6} f_p\int \frac{d k_0^{(2)}}{k_0^{(2)}}\int d^2 k_\perp \frac{8 p_0^2 m_V^4}{(k_\perp^2+m_V^2)^2 k_\perp^2} \frac{k_\perp^2+m_V^2}{ p_0 x}\nn
&=&\int\frac{d^3 p}{ p_0 (2\pi)^6}f_p\int \frac{d x}{x^2}\int \frac{d^2 k_\perp}{k_\perp^2} \frac{ m_V^4}{(k_\perp^2+m_V^2) }\nn
&\simeq&\int\frac{d^3 p}{ p_0 (2\pi)^6}f_p \pi m_V^2 \log(m_V^2/(eT)^2)\times\l[\int \frac{dx}{x^2}= \frac{p_0}{m_V}\r]\nn
&\simeq&\int \frac{d^3 p}{(2\pi)^3 p_0}f_p  \l[\frac{m_V p_0}{8\pi^2}\log (m_V^2/(eT)^2) \r].
\eea
Let us make a few comments regarding this expression. We can see that in the wall frame the pressure is proportional to 
$\Delta \mathcal{P}\propto \int d^3 p f_p$, however $d^3 p$ is not invariant under the boost and in the plasma frame it will lead to the additional $\gamma$ factor
\bea
\Delta  \mathcal{P}\propto \gamma T^3 m_V.
\eea
Another important point we would  like to stress is that the minimal value of the transverse momenta is  cut in the IR at the scale  $k_\perp^{\text{min}} \sim e T $, due to the screening of the long wavelength  modes by the temperature effects ($e$ is the gauge coupling). 
We can see that the pressure  is dominated by the emission of the \emph{soft} photons, which provides the $\gamma$ enhancement.
{ Note that this transition is dominated by the transverse momenta  
$k_\perp\sim m_V$, so that momentum transfer always satisfies $\Delta p_z L\sim \frac{m_V^2}{p_0 x}L\sim \frac{m_V}{p_0 x}\lesssim 1$. Thus the finite width effect effect will not lead to additional suppression of the transition.}
 In the next subsection we will rederive the same result using semi-classical equivalent photon approximation.

\subsection{Equivalent photon approximation}
It is well-known that the effect of the soft and collinear photons can be taken into account using the equivalent photon approximation (EPA)\cite{fermi:24,vonWeizsacker:1934nji,Williams:1934ad,landau:34} (see for  review \cite{Peskin:1995ev,Akhiezer:1986yqm,Berestetsky:1982aq}). In other words, an initial fermion state can be thought as a state made of photons and fermions with the photons  distributed according to the Weizsacker-Williams parton distribution function 
\bea
f_\gamma(x) =\frac{e^2}{8\pi^2}\log \frac{m_V^2}{(eT)^2}\l[\frac{1+(1-x)^2}{x}\r],
\eea 
where we are using the information from Eq.\eqref{eq:pres12} that the pressure is dominated by the $k_\perp \lesssim m_V$ and the minimal value of transverse momenta scales as $\sim eT$\footnote{One can also argue that $k_\perp \lesssim m_V$ by noting that in the limit $px , k_\perp \gg m_V$ four momentum is approximately conserved, which should strongly suppress the splitting.  }. We also know that a photon with a phase-dependent mass going through the wall will lose (deposited in the wall) z-momenta, of the order $\Delta p_z\sim \frac{m_V^2}{2 E_{\gamma}}$.
Then the pressure can be trivially estimated to be
\bea
\label{eq:splitting}
\mathcal{P}_{1\to 2}^{eq. \gamma}&=&\underbrace{\int \frac{d^3 p}{(2\pi)^3 }f_p}_{\text{incident fermions}} \int_{m_V /p}^1 dx f_\gamma(x) \times \underbrace{\frac{m_V^2}{2 p x}}_{\text{momentum transfer}}\nn
&=&\int \frac{d^3 p}{(2\pi)^3 }f_p\times \l[\frac{e^2}{8\pi^2}  m_V\r]\log \frac{m_V^2}{e^2 T^2}
\eea
which leads to the exactly same result as the expression in Eq.\eqref{eq:pres12}. Intuitively the $\gamma$ factor in the pressure comes from the two following  effects: both the photon distribution function as well as momentum transfer to the wall are  enhanced by the factor  $1/x$, which together allows to enhance the pressure by the additional factor $p_0/m_V\sim \gamma$. One may wonder what will be the effect  of the particles which do not have enough energy to pass through the wall, since for them the photon  distribution function will be even larger. However,  in that case the momentum transfer to the wall will  scale as $p_0 x$, so that the pressure will scale as
\bea
\mathcal{P}_{1\to 2}^{\rm reflection}&=&\int \frac{d^3 p}{(2\pi)^3 }f_p \int dx f_\gamma(x) \times 2 p_0 x\nn
&=&\int \frac{d^3 p}{(2\pi)^3 }f_p\times \l[\frac{e^2}{2\pi^2}  (x_{\text{max}}-x_{\text{min}}) p_0\r]\log \frac{m_V^2}{e^2T^2}
\eea
\bea
x_{\text{min}}\sim k_\perp /p_0\sim T/p_0,~~~~x_{\text{max}}\sim m_V/p_0
\eea
leading to the pressure from reflection
\bea
\mathcal{P}_{1\to 2}^{\rm reflection} \simeq \int \frac{d^3 p}{(2\pi)^3 }f_p \l[\frac{e^2}{2\pi^2} m_V \log \frac{m_V^2}{e^2T^2}\r].
\label{eq:refelection}
\eea
We have again the friction effect growing with the Lorentz factor $\gamma$. However, note that our  calculation  becomes  questionable in this regime, since we need   $p_0 x L\gg 1$  in order to remain in the WKB validity range.

We can generalize the Eq.\eqref{eq:splitting} for arbitrary splitting and the resulting pressure will be
\bea
\label{eq:splittinggen}
\mathcal{P}_{A\to B C }=\int \frac{d^3 p }{(2\pi)^3}f_p\int^1_{m_B/p} d x \frac{m^2_B}{2 px}\frac{\alpha}{2\pi}\log \frac{m_B^2}{e^2 T^2} P_{B\leftarrow A}(x)
\eea
where $B$ is the soft particle and $P_{B\leftarrow A}(x)$ are Altarelli-Parisi \cite{Altarelli:1977zs} splitting functions. Then it is obvious that a friction  proportional to $\propto \gamma $ can appear only from the splitting when the splitting functions scale as $1/x$ for small values of $x$. This is the case only when the soft final state is a vector boson, which confirms the results of \cite{Bodeker:2017cim}.

The expression of the pressure in Eq.\eqref{eq:splittinggen} was derived assuming single soft vector boson emission, and it corresponds to the solution of the DGLAP equations \cite{Altarelli:1977zs,Gribov:1972rt,Dokshitzer:1977sg}
\bea
\frac{d f_B(x,Q)}{d \log Q}=\frac{\alpha}{\pi}\int_x^1 \frac{d z}{z}P_{B\leftarrow A}(z)f_A \l(\frac{x}{z},Q\r),
\eea
up to the order $\mathcal{O}(\alpha^2)$ starting with initial conditions at the scale $Q= e T$
\bea
f_B(x,eT)=0,~~~f_A(x,e T)=\delta(x-1) . 
\eea
The multiple emissions can be taken into account by solving the system of the DGLAP equations, however these will lead to only higher order in $\mathcal{O}(\alpha \log \frac{m_V}{e T})$ corrections. We observe that in absence of log-enhancement $\log \frac{m_V}{e T} \sim \mathcal{O}(1)$, any multiple emission will be suppressed by powers of the coupling $\alpha$ with respect to the single-emission result. 

\subsection{Another calculation of the NLO  friction effects }
 Recently there was another calculation \cite{Hoeche:2020rsg}
%\footnote{We acknowledge J. Turner and A. Long for discussions on the results of Ref. \cite{Hoeche:2020rsg}.}
of the friction which tried to take into account effects of the soft emission.
The resulting friction pressure for the fermion emitting soft vector bosons was found to scale as 
\bea
\label{2007.10343}
\mathcal{P}^{\cite{Hoeche:2020rsg}}\sim \alpha \gamma^2 T^4.
\eea
However note  that Eq. \eqref{2007.10343} does not have the correct $m_\psi,m_V\to 0$ limit (vanishing masses of the fermion and the vector boson). Indeed in the case when both $m_\psi,m_V=0$ the particles do not interact with the wall and it becomes completely transparent. However the particles which do not interact with the bubble wall cannot induce any friction so that $\mathcal{P}^{\rm friction}|_{m_\psi,m_V\to 0}\to 0$.  This signals the inconsistency of the Eq. \eqref{2007.10343}. On general grounds the inconsistency of Eq. \eqref{2007.10343}
 can be seen directly by noting that there is no dependence  on the order parameter differentiating two phases separated by the bubble wall.

\section{Summary}
\label{sec:conc}
In summary, we recapitulate the main results of this paper. We have studied the friction forces acting on the relativistically expanding bubble at leading and next-to-leading order in the coupling $\alpha$. We have shown that generically new heavy particles, even if they are completely decoupled at the scale of the phase transition, can provide a significant contribution to the friction force. This  effect can significantly modify the dynamics of the bubble wall expansion and in particular it can prevent the runaway behaviour of the bubble expansion, which results in different stochastic gravitational backgrounds. We have illustrated the effect using a toy model example where we show that new states  { being $\lesssim10^5$ heavier than the scale of the symmetry breaking},  can be active source of friction and prevent the infinite acceleration of the bubbles. 

Beside this new result we have reviewed the NLO friction results of the Ref.\cite{Bodeker:2017cim}, where it was shown that the soft vector boson emission leads to a new component of the friction pressure which scales proportionally to $\propto \gamma$.
We have presented an alternative derivation of this effect  using the equivalent photon approximation, which provides an intuitive picture of the origin of the friction $\propto \gamma$ as well as commented on the importance of the higher order effects and the ways to include them in the calculation.

\section*{Acknowledgments}
This work was in part supported by the MIUR contract 2017L5W2PT. We thank as well A. Long and J. Turner for useful discussions concerning the results of \cite{Hoeche:2020rsg}.

\appendix

\section{Transition pressure }
\label{sec:pressuredetails}
In this appendix we will review, for the sake of completeness, the pressure from transition of the particles through the wall. We will focus on the limit $
\gamma\gg 1$ and  will always assume that the mean free path of the particles is larger 
than the wall width. In this case we can treat particles quasi-classically and consider only individual 
interactions with the wall.
This discussion is not new and was already presented in the papers \cite{Dine:1992wr,Arnold:1993wc} and recently reviewed in \cite{Mancha:2020fzw}(where the analytical results for the pressure have been reported).
The particle will follow the usual thermal distribution, which in the frame of the wall becomes
\bea
f(E,p,T)=f\l(\frac{p_\mu u^\mu}{T}\r)=f\l(\frac{\gamma(E+v p_z)}{T}\r), 
\eea
where we have assumed like in Fig.\ref{fig:cartoon} that the wall moves along the positive $z$ direction with velocity $v$.
We will assume that the particle is incident on the wall with mass $m_1$ and on the other side it has mass $m_2$. 
The pressure on the wall is originating from the following three processes (we follow closely the notations of  \cite{Dine:1992wr}).

\bit
\item  Reflection from the wall, when the incident particle does not have enough momentum or energy to pass through the wall:
 \bea
\Delta \pr ^{r}& =&\frac{2}{4\pi^2}\int_{m_1}^{m_2}  d E \int_{-\sqrt{E^2 -m_1^2}}^{0} d p_z\l[ p_z^2 f\l(\frac{\gamma(E+v p_z)}{T}\r)\r]\nn
&+&\frac{2}{4\pi^2}\int_{m_2}^{\infty}  d E \int_{-\sqrt{m_2^2 -m_1^2}}^{0} d p_z\l[ p_z^2 f\l(\frac{\gamma(E+v p_z)}{T}\r)\r]
\label{Pres:ref}
\eea 
in this case  the momentum transfer  to the wall is $\Delta p_z = 2p_z$.
\item  Transition through the wall, the pressure is generated due to the change of  momenta of the particle with $\Delta p_z =  p_z+\sqrt{p_z^2-(m_2^2-m_1^2)}$:
\bea
\Delta \pr^{t+}& =&\frac{1}{4\pi^2}\int_{m_2}^\infty  d E \int_{-\sqrt{E^2 -m_1^2}}^{-\sqrt{m_2^2-m_1^2}} d p_z\l[p_z (p_z+\sqrt{p_z^2-(m_2^2-m_1^2)})f\l(\frac{\gamma(E+v p_z)}{T}\r)\r]\nn
\label{Pres:+}
\eea 
\item Transition in the opposite direction with $\Delta p_z = \sqrt{p_z^2+m_2^2-m_1^2}-p_z$:
\bea
\Delta \pr^{t-}& =&\frac{1}{4\pi^2}\int_{m_2}^\infty  d E  \int_{0}^{\sqrt{E^2-m_2^2}} d p_z\l[p_z (\sqrt{p_z^2+m_2^2-m_1^2}-p_z)f\l(\frac{\gamma(E+v p_z)}{T}\r)\r]\nn
\label{Pres:-}
\eea
\eit
Let us start by considering the transition pressure Eq.\eqref{Pres:+}. 
Introducing the new variables
\bea
&& Y\equiv \frac{\gamma(E+v p_z)}{T},~~~k\equiv -\frac{p_z}{T},
\eea
the expression for the pressure becomes:
\bea
\Delta \pr^{t+}  &=&\frac{T^4}{4\pi^2 \gamma}  \int_{\sqrt{m_2^2-m_1^2}/T}^\infty d k k(k-\sqrt{k^2-(m_2^2-m_1^2)/T^2})\int_{\gamma ({{\sqrt{k^2+m_1^2/T^2}}}-v k)}^\infty f(Y) dY\nn
&=&-\frac{T^2 (m_2^2-m_1^2)}{8\pi^2\gamma}\int_{\sqrt{m_2^2-m_1^2}/T}^\infty d k \int_{\gamma({{\sqrt{k^2+m_1^2/T^2}}}-v k)}^\infty f(Y) dY,
\label{eq:varred}
\eea
where we have expanded the momentum difference $(k-\sqrt{k^2-(m_2^2-m_1^2)/T^2})$ in the large $k $ limit. We can see that the integral is non-vanishing if the lower limit of the second integral  is small
\bea
\l\{ \gamma (\sqrt{k^2+m^2_1/T^2}  - v k)\r\}  \sim \frac{k }{2\gamma}+\frac{m_1^2}{2 T^2}\frac{\gamma}{k} \lesssim O(1)\Rightarrow
m_1\lesssim  T .
\eea
Otherwise the pressure effects will be strongly suppressed by the Boltzmann factor $\exp[-\frac{m_1}{T}]$, which is obvious, since  the energy of the massive particle is always larger than its mass.  On top of this, looking at  the lower limit of the $k$ integration is  we can conclude that 
\bea
Y\lesssim 1\Rightarrow \sqrt{m_2^2-m_1^2}\lesssim  \gamma T,
\eea
which is just the necessary  condition for the particle to pass through the wall. Combining these two conditions we observe that the friction is efficient if only
\bea
m_2 < \gamma T,~~~ m_1 <T.
\eea
Performing the integration we will obtain for the friction
\bea
\Delta \pr^{t+}|_{\gamma T/m_2\to {{\infty}}}=\frac{m_2^2-m_1^2}{24} T^2 .
\eea
Using a similar analysis we can argue that the reflection pressure and the transmission from the opposite side are vanishingly  small in $\gamma\to \infty $ limit. Indeed  setting $m_1\to 0$ for simplicity and using the same variable redefinition as in Eq. \eqref{eq:varred} we will get
\bea
\Delta \pr ^{r}&=&I_1+I_2\nn
I_1&=&\frac{2}{4\pi^2}\int_{0}^{m_2}  d E \int_{-\sqrt{E^2 -m_1^2}}^{0} d p_z\l[ p_z^2 f\l(\frac{\gamma(E+v p_z)}{T}\r)\r]\nn
&=&\frac{T^4}{2\pi^2 \gamma}\int_0^{m_2/T} d k k^2\int_\frac{k}{2\gamma}^{\gamma(m_2/T-k)}  dY f(Y)\propto \gamma^{-1}\to 0 \nn
I_2&=&\frac{T^4}{2\pi^2 \gamma}\int _0^{m_2/T}dk k^2 \int_{\gamma(m_2/T-v k)}^\infty f(Y) dY \propto \gamma^{-2}\to 0.
\eea
At last the pressure from the transition in the opposite direction
$\Delta \pr^{t-}$ is always suppressed since the argument of the distribution function  is always larger than  one  $\sim  \frac{\gamma m}{T}\gg1$ .
\begin{figure}
\includegraphics[scale=0.5]{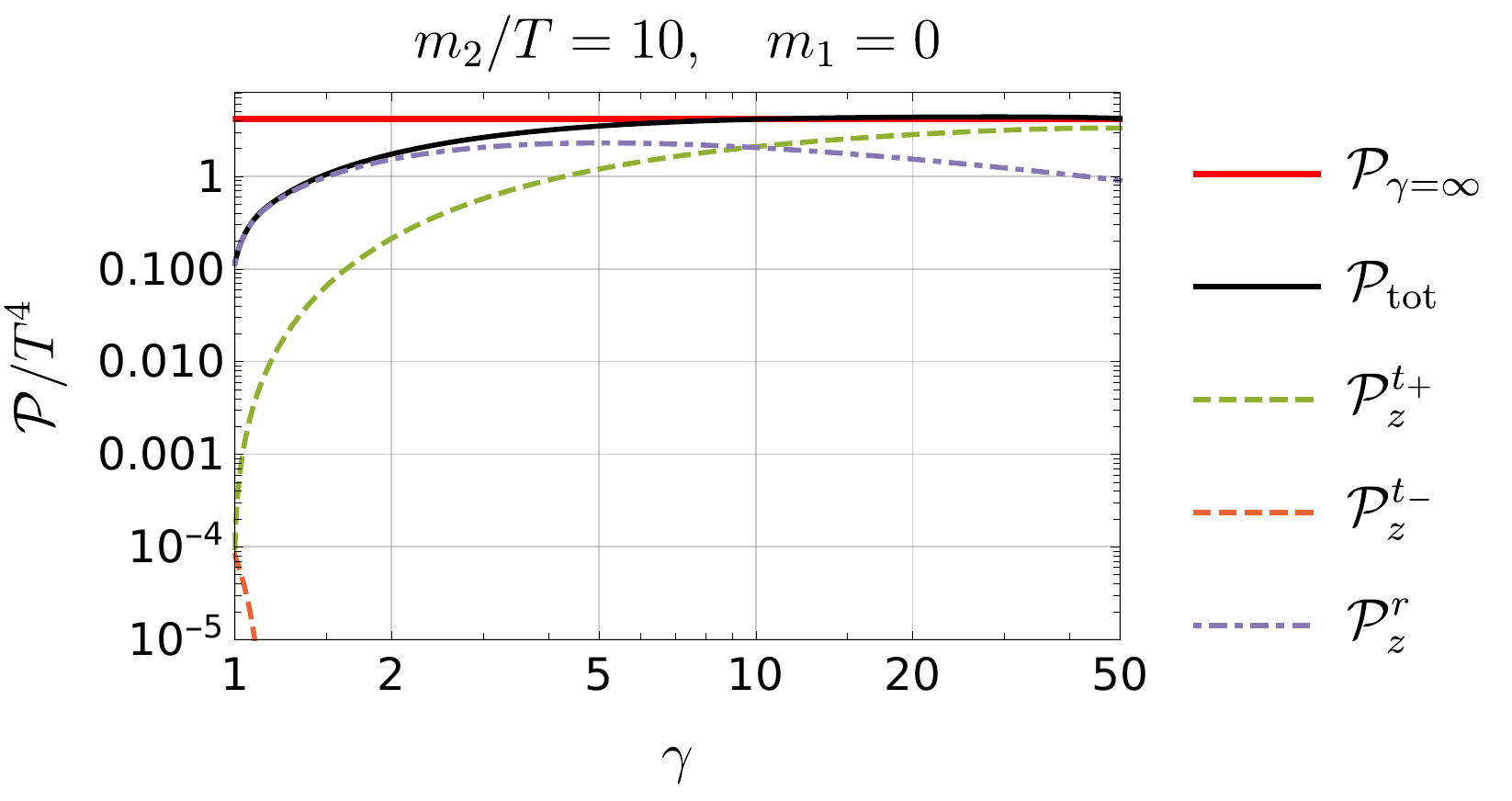}
\includegraphics[scale=0.5]{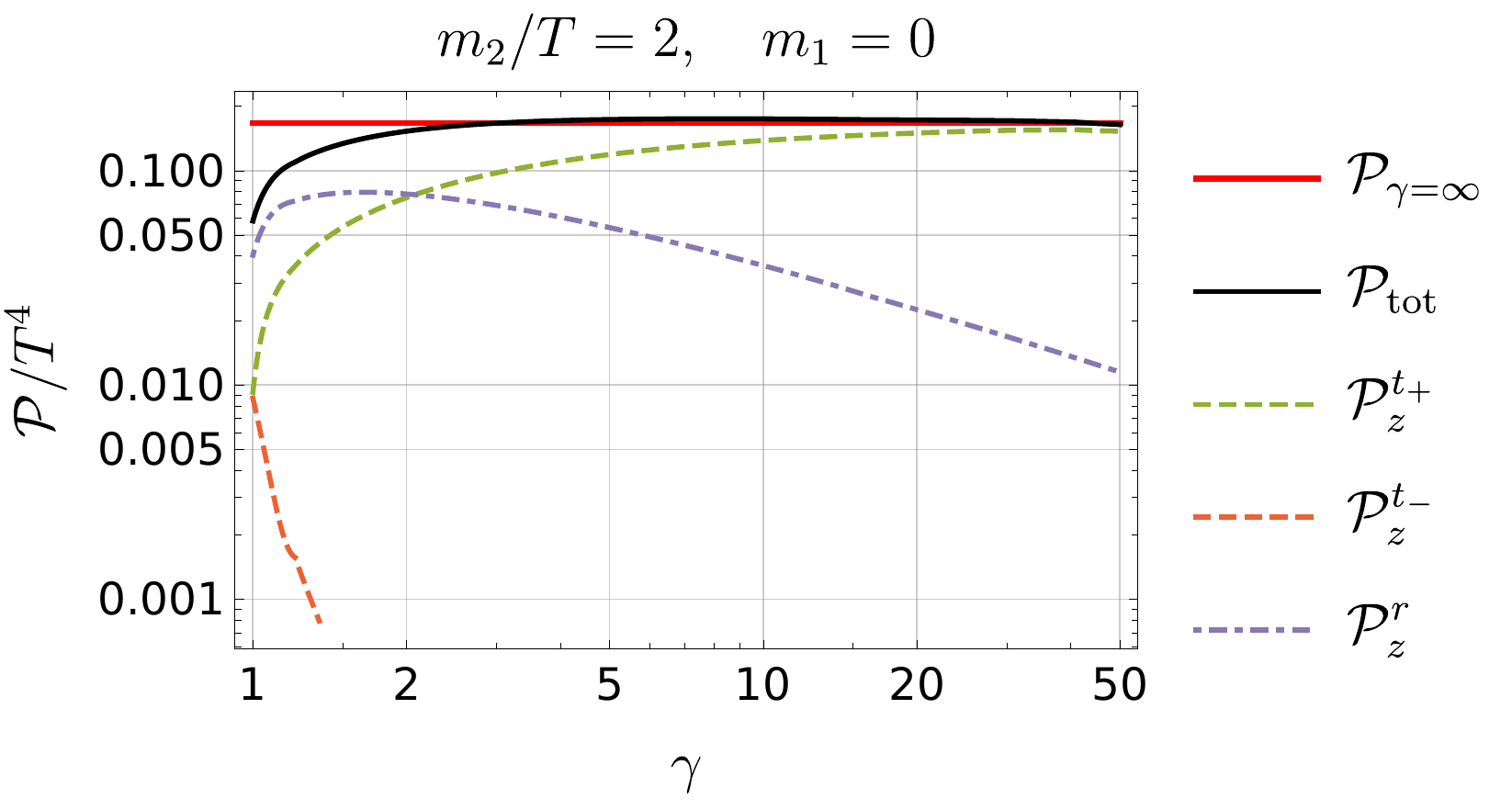}
\caption{Illustration of the forward transmission pressure, the reflection pressure, the total pressure and the LO order approximation. $\frac{m_2}{T} = 10, 2$ respectively on the Left and the Right.}
\label{Limitgamma}
\end{figure}
We confirm these estimates using our numerical calculation illustrated on the Figure \ref{Limitgamma}, where we plotted  the various contributions to the total pressure. For the various contributions to the pressure the following approximate relations are true in the mass range $\frac{m}{T}\sim 1-10$:
\bea
 \mathcal{P}^{r}_z \approx \mathcal{P}^{t_+}_z \approx 0.4\times \mathcal{P}_{\gamma \to \infty} \text{    for   } \gamma T = m_0
\\\mathcal{P}^{t_-}_z \approx \mathcal{P}^{r}_z \approx 0, \quad \mathcal{P}^{t_+}_z \approx 0.9\times \mathcal{P}_{\gamma \to \infty} \text{    for   } \gamma T = 10 m_0 .
\eea

%\begin{figure}
%	\centering
%	\includegraphics[scale=0.5]{presM0m21T1.pdf}
%	\includegraphics[scale=0.5]{presM0m22T1.pdf}
%	\includegraphics[scale=0.5]{presM0m205T1.pdf}
%	\caption{Illustration of each contribution of the pressure as a function of $\gamma$, for small velocities. We observe that when the wall is very slow, the reflection is dominant and the two types of transmission give comparable contributions. As the wall accelerates the ``backward transmission'' and the reflection are damped while the ``forward transmission'' quickly increases. The moment when ``forward transmission'' takes over is earlier as we take larger $\frac{m_0}{T}$.  {\bf do we want to keep this plot?  since for such small $\gamma$ it is not clear that our method is valid}  }
%	\label{pressurem}
%	\end{figure}

\section{Examples of the friction induced by the heavy particles}
\label{sec:frictionheavy}
In Section \ref{sec:mixing}, we have shown that the mixing of a light and a heavy fermion can lead to the friction, which we called \emph{mixing pressure}. We can find a similar effect  in the theories with  scalars  fields only. In this appendix we will present two such examples of the non-vanishing pressure from the heavy fields. 
\subsection{Pressure from scalar mixing}
Let us start by considering the following model:
\bea
{\cal L}=\frac{1}{2}(\d s)^2+ \frac{1}{2}(\d \phi)^2- B s^2 \phi -\frac{ M^2\phi^2}{2},
\eea
where the phase transition occurs along the $s$ field direction and  there is a hierarchy between the VEV of the $s$ field and the mass of the $\phi$, $M\gg \langle s\rangle $.
In this case, following the lines of Section \ref{sec:mixing} the mixing between the $s$ field and the heavy mass eigenstate inside the wall will scales as
\bea
\theta_{s- \phi}\sim \frac{2 B \langle s\rangle  }{2M^2}.
\eea
 This mixing leads to a pressure of the form
\bea
\mathcal{P}_{\text{mixing}}\sim T^2 \frac{B^2 \langle s\rangle^2}{24M^2}  \theta (\gamma T -M^2L). 
\eea 
Note that  the friction is suppressed by a factor $\frac{B^2}{M^2}$ with respect to the pressure induced by fermionic mixing. This suppression disappears in the limit $B\to M$, which is the maximal value of $B$ allowed by the technical naturalness arguments.

\subsection{Pressure from scalar $1\to 2$ splitting}

Another example of  friction from heavy particles effect can be observed in the following model:
\bea
{\cal L}=\frac{(\d s)^2}{2}+\frac{(\d \phi)^2}{2}- V(s)- \frac{M^2 \phi^2}{2}- \lambda \phi^2 s^2 ,
\label{Model2_heavy}
\eea
where again we will be interested in the limit $M\gg \langle s \rangle$.
We will consider the process $s\to \phi \phi$, where $s$ is a field getting a VEV. The pressure will be now sustained by a $s \to \phi\phi$ decay in the wall. Following the procedure outlined in the section \ref{sec:nlo} we can work out the matrix element:
\bea
&&A=-\frac{k_\perp^2}{x(1-x)}+m_s^2-\frac{M^2}{1-x}-\frac{M^2}{x}\nn
&&V_h=\lambda \langle s\rangle  \qquad V_s = 0,\nn 
&& \Rightarrow |\mathcal{M}|^2=\frac{4 p_0^2 \lambda^2 \langle s\rangle^2}{|A|^2}\sim  \frac{4 p_0^2 \lambda^2 \langle s\rangle^2 x^2(1-x)^2}{(k_\perp^2+M^2)^2}, 
\eea
we used $s(h)$ subscripts to denote the components on the symmetric side (on the broken side). 
As already emphasized in Section \ref{sec;mixing_more_details}, this computation is too naïve and ignores the width of the wall. {  As in Section \ref{sec;mixing_more_details}, to illustrate the effect of the non-negligible width of the wall, we consider a linear profile. Using the WKB phases and the notations of \cite{Bodeker:2017cim}, the $\mathcal{M}$ matrix in this case writes 
  \bea
  \mathcal{M} = \int  \limits_{-\infty}^{\infty}dz e^{i \int\limits_0^{z} p^{s}_z(z') dz'}e^{-iq^{\phi}_z z}e^{-i k^{\phi}_z z} V(z)
  \eea
  with $ p^{s}_z(z) = \sqrt{E^2 - k^2_\perp - (z/L)^2 m_s^2} \approx E$ the momemtum of the incoming $s$ particle and $k^{\phi}_z, q^{\phi}_z$ the momentum of the two $\phi$ outgoing particles. The matrix element becomes
   \bea
  \mathcal{M} = \int\limits_{-\infty}^{\infty} dz e^{i p^{s}_z z}e^{-iq^{\phi}_z z}e^{-i k^{\phi}_z z} V(z) \equiv \int\limits_{-\infty}^{\infty} dz e^{i \Delta p_z z} V(z),
  \eea
  where we defined $\Delta p_z \equiv p^{s}_z - q^{\phi}_z - k^{\phi}_z \approx \frac{M^2}{2E}$, the momentum exchange.
  The integral along the wall direction naturally splits into three parts
   \begin{equation}
 \mathcal{M} = \mathcal{M}_{\text{before wall}}+ \mathcal{M}_{\text{inside wall}}+ \mathcal{M}_{\text{after wall}}
  \end{equation}
  The first term is zero, as the interaction is forbidden on the symmetric side, the second term is given by $ (1- e^{i \Delta p_z L}) \frac{V_h}{i\Delta p_z^2 L} + V_h\frac{e^{i\Delta p_z L}}{i\Delta p_z} $ and the third term by $\frac{V_h}{i\Delta p_z} (-e^{i\Delta p_z L} + e^{i \infty}) $.
  Putting together the relevant pieces, we obtain
  \begin{equation}
 \mathcal{M} = V_h\frac{1 - e^{i\Delta p_z L}}{L \Delta p_z^2}.
  \end{equation}
  The final matrix element is 
  \begin{equation}
 |\mathcal{M}|^2 = \mathcal{M} \mathcal{M}^{\star} = \frac{V^2}{\Delta p_z^2} \bigg(\frac{\sin \alpha}{\alpha} \bigg)^2, \qquad \alpha \equiv \frac{L \Delta p_z}{2}.
  \end{equation}
   We observe again that to account for the width of the wall, we need to introduce the suppression factor $\big(\frac{\sin \alpha}{\alpha} \big)^2$. 
Then the pressure will be:
\bea
\mathcal{P}_{1 \to 2}&\simeq &\int \frac{d^3 p}{(2\pi)^3p_0^2}f_p\int \frac{dx}{32\pi^2 x(1-x)}\int d k_\perp^2 \frac{4 p_0^2 \lambda^2 \langle s\rangle^2 x^2(1-x)^2}{(k_\perp^2+M^2)^2} 
\\ \nonumber
&\times &\l[\frac{k_\perp^2+M^2}{2p_0 x(1-x)}\r] \times \bigg(\frac{\sin \alpha}{\alpha} \bigg)^2 \theta(p_0-2M)\nn
&= & \int \frac{d^3 p}{(2\pi)^3 p_0}f_p{\frac{\lambda^2}{16\pi^2} \langle s\rangle^2}\int dx \int  \frac{ d k_\perp^2}{k_\perp^2+M^2} \times \bigg(\frac{\sin \alpha}{\alpha} \bigg)^2\theta(p_0-2M)
\label{friction_heavy}
\eea
where the $\theta$ function appears from the trivial requirement that we need enough energy to produce the two heavy states. Thus the pressure becomes
\bea
\mathcal{P}_{1 \to 2}&\simeq& \int \frac{d^3 p}{(2\pi)^3p_0}f_p\frac{\lambda^2}{16\pi^2} \langle s\rangle^2\times \bigg(\frac{\sin \alpha}{\alpha} \bigg)^2\theta (p_0-2M)
\\
&\simeq & \frac{\lambda^2}{16\pi^2}\frac{\langle s\rangle^2}{24} T^2 \times \bigg(\frac{\sin \alpha}{\alpha} \bigg)^2\theta (\gamma T-2M).
\\
&\simeq &
\frac{\lambda^2}{16\pi^2}\frac{\langle s\rangle^2}{24} T^2 \times \theta(\gamma T - M^2L)\theta (\gamma T-2M).
\eea
}
 So again, in the limit of small exchange momentum, the friction is not suppressed by the large mass of the field $\phi$.

\bibliographystyle{JHEP}
{\footnotesize
\bibliography{biblio}}

\providecommand{\href}[2]{#2}\begingroup\raggedright\begin{thebibliography}{10}

\bibitem{Witten:1984rs}
E.~Witten {\em Phys. Rev.} {\bf D30} (1984) 272--285.

\bibitem{Kuzmin:1985mm}
V.~Kuzmin, V.~Rubakov, and M.~Shaposhnikov {\em Phys. Lett. B} {\bf 155} (1985)
  36.

\bibitem{Shaposhnikov:1986jp}
M.~Shaposhnikov {\em JETP Lett.} {\bf 44} (1986) 465--468.

\bibitem{Grasso:2000wj}
D.~Grasso and H.~R. Rubinstein {\em Phys. Rept.} {\bf 348} (2001) 163--266,
  [\href{http://arxiv.org/abs/astro-ph/0009061}{{\tt astro-ph/0009061}}].

\bibitem{Liu:1992tn}
B.-H. Liu, L.~D. McLerran, and N.~Turok {\em Phys. Rev. D} {\bf 46} (1992)
  2668--2688.

\bibitem{Dorsch:2018pat}
G.~C. Dorsch, S.~J. Huber, and T.~Konstandin {\em JCAP} {\bf 1812} (2018),
  no.~12 034, [\href{http://arxiv.org/abs/1809.04907}{{\tt arXiv:1809.04907}}].

\bibitem{Konstandin:2014zta}
T.~Konstandin, G.~Nardini, and I.~Rues {\em JCAP} {\bf 1409} (2014), no.~09
  028, [\href{http://arxiv.org/abs/1407.3132}{{\tt arXiv:1407.3132}}].

\bibitem{Moore:1995ua}
G.~D. Moore and T.~Prokopec {\em Phys. Rev. Lett.} {\bf 75} (1995) 777--780,
  [\href{http://arxiv.org/abs/hep-ph/9503296}{{\tt hep-ph/9503296}}].

\bibitem{Moore:1995si}
G.~D. Moore and T.~Prokopec {\em Phys. Rev.} {\bf D52} (1995) 7182--7204,
  [\href{http://arxiv.org/abs/hep-ph/9506475}{{\tt hep-ph/9506475}}].

\bibitem{Laurent:2020gpg}
B.~Laurent and J.~M. Cline {\em Phys. Rev. D} {\bf 102} (2020), no.~6 063516,
  [\href{http://arxiv.org/abs/2007.10935}{{\tt arXiv:2007.10935}}].

\bibitem{Bodeker:2009qy}
D.~Bodeker and G.~D. Moore {\em JCAP} {\bf 0905} (2009) 009,
  [\href{http://arxiv.org/abs/0903.4099}{{\tt arXiv:0903.4099}}].

\bibitem{Bodeker:2017cim}
D.~Bodeker and G.~D. Moore {\em JCAP} {\bf 1705} (2017), no.~05 025,
  [\href{http://arxiv.org/abs/1703.08215}{{\tt arXiv:1703.08215}}].

\bibitem{Dine:1992wr}
M.~Dine, R.~G. Leigh, P.~Y. Huet, A.~D. Linde, and D.~A. Linde {\em Phys. Rev.}
  {\bf D46} (1992) 550--571, [\href{http://arxiv.org/abs/hep-ph/9203203}{{\tt
  hep-ph/9203203}}].

\bibitem{Arnold:1993wc}
P.~B. Arnold {\em Phys. Rev. D} {\bf 48} (1993) 1539--1545,
  [\href{http://arxiv.org/abs/hep-ph/9302258}{{\tt hep-ph/9302258}}].

\bibitem{Mancha:2020fzw}
M.~Barroso~Mancha, T.~Prokopec, and B.~Swiezewska
  \href{http://arxiv.org/abs/2005.10875}{{\tt arXiv:2005.10875}}.

\bibitem{fermi:24}
E.~Fermi {\em Zeitschrift fur Physik} {\bf 29} (1924), no.~1 315--327.

\bibitem{vonWeizsacker:1934nji}
C.~von Weizsacker {\em Z. Phys.} {\bf 88} (1934) 612--625.

\bibitem{Williams:1934ad}
E.~Williams {\em Phys. Rev.} {\bf 45} (1934) 729--730.

\bibitem{landau:34}
L.~Landau and E.~Lifshitz {\em Phys.Z.Sowjetunion} {\bf 6} (1934), no.~1 612.

\bibitem{Balaji:2020yrx}
S.~Balaji, M.~Spannowsky, and C.~Tamarit
  \href{http://arxiv.org/abs/2010.08013}{{\tt arXiv:2010.08013}}.

\bibitem{Mikheev:1986if}
S.~Mikheev and A.~Smirnov {\em Sov. Phys. JETP} {\bf 64} (1986) 4--7,
  [\href{http://arxiv.org/abs/0706.0454}{{\tt arXiv:0706.0454}}].

\bibitem{Bilenky:2010zza}
S.~Bilenky, {\em {Introduction to the physics of massive and mixed neutrinos}},
  vol.~817.
\newblock 2010.

\bibitem{Ellis:2018mja}
J.~Ellis, M.~Lewicki, and J.~M. No \href{http://arxiv.org/abs/1809.08242}{{\tt
  arXiv:1809.08242}}. [JCAP1904,003(2019)].

\bibitem{Ellis:2019oqb}
J.~Ellis, M.~Lewicki, J.~M. No, and V.~Vaskonen {\em JCAP} {\bf 1906} (2019),
  no.~06 024, [\href{http://arxiv.org/abs/1903.09642}{{\tt arXiv:1903.09642}}].

\bibitem{Espinosa:2010hh}
J.~R. Espinosa, T.~Konstandin, J.~M. No, and G.~Servant {\em JCAP} {\bf 1006}
  (2010) 028, [\href{http://arxiv.org/abs/1004.4187}{{\tt arXiv:1004.4187}}].

\bibitem{Caprini:2019egz}
C.~Caprini et~al. \href{http://arxiv.org/abs/1910.13125}{{\tt
  arXiv:1910.13125}}.

\bibitem{Breitbach:2018ddu}
M.~Breitbach, J.~Kopp, E.~Madge, T.~Opferkuch, and P.~Schwaller {\em JCAP} {\bf
  1907} (2019), no.~07 007, [\href{http://arxiv.org/abs/1811.11175}{{\tt
  arXiv:1811.11175}}].

\bibitem{Azatov:2019png}
A.~Azatov, D.~Barducci, and F.~Sgarlata {\em JCAP} {\bf 07} (2020) 027,
  [\href{http://arxiv.org/abs/1910.01124}{{\tt arXiv:1910.01124}}].

\bibitem{Weinberg:1973am}
E.~J. Weinberg, {\em {Radiative corrections as the origin of spontaneous
  symmetry breaking}}.
\newblock PhD thesis, Harvard U., 1973.
\newblock \href{http://arxiv.org/abs/hep-th/0507214}{{\tt hep-th/0507214}}.

\bibitem{Curtin:2016urg}
D.~Curtin, P.~Meade, and H.~Ramani {\em Eur. Phys. J.} {\bf C78} (2018), no.~9
  787, [\href{http://arxiv.org/abs/1612.00466}{{\tt arXiv:1612.00466}}].

\bibitem{Coleman:1977py}
S.~R. Coleman {\em Phys. Rev.} {\bf D15} (1977) 2929--2936. [Erratum: Phys.
  Rev.D16,1248(1977)].

\bibitem{Linde:1980tt}
A.~D. Linde {\em Phys. Lett.} {\bf 100B} (1981) 37--40.

\bibitem{Linde:1981zj}
A.~D. Linde {\em Nucl. Phys.} {\bf B216} (1983) 421. [Erratum: Nucl.
  Phys.B223,544(1983)].

\bibitem{Enqvist:1991xw}
K.~Enqvist, J.~Ignatius, K.~Kajantie, and K.~Rummukainen {\em Phys. Rev.} {\bf
  D45} (1992) 3415--3428.

\bibitem{Peskin:1995ev}
M.~E. Peskin and D.~V. Schroeder, {\em {An Introduction to quantum field
  theory}}.
\newblock Addison-Wesley, Reading, USA, 1995.

\bibitem{Akhiezer:1986yqm}
A.~Akhiezer and V.~Berestetskii, {\em {Quantum electrodynamics}}.
\newblock Interscience Publishers, New York, 9, 1986.

\bibitem{Berestetsky:1982aq}
V.~Berestetskii, E.~Lifshitz, and L.~Pitaevskii, {\em {QUANTUM
  ELECTRODYNAMICS}}, vol.~4 of {\em Course of Theoretical Physics}.
\newblock Pergamon Press, Oxford, 1982.

\bibitem{Altarelli:1977zs}
G.~Altarelli and G.~Parisi {\em Nucl. Phys. B} {\bf 126} (1977) 298--318.

\bibitem{Gribov:1972rt}
V.~Gribov and L.~Lipatov {\em Sov. J. Nucl. Phys.} {\bf 15} (1972) 675--684.

\bibitem{Dokshitzer:1977sg}
Y.~L. Dokshitzer {\em Sov. Phys. JETP} {\bf 46} (1977) 641--653.

\bibitem{Hoeche:2020rsg}
S.~H\"oche, J.~Kozaczuk, A.~J. Long, J.~Turner, and Y.~Wang
  \href{http://arxiv.org/abs/2007.10343}{{\tt arXiv:2007.10343}}.

\end{thebibliography}\endgroup
\end{document}